\documentclass[utf8]{forarxiv} 

\usepackage{url,hyperref,lineno,microtype,subcaption}

\usepackage{amsmath,amssymb,amsfonts}
\usepackage{graphicx}
\usepackage{xcolor}

\usepackage{float} 
\usepackage{subcaption}
\usepackage{tabularx}

\def\firstAuthorLast{Morita {et~al.}} %use et al only if is more than 1 author
\def\Authors{Junya Morita\,$^{1,*}$, Thanakit Pitakchokchai\,$^{1,*}$, Giri Basanta Raj\,$^{1,*}$, Yusuke Yamamoto\,$^{1}$ Hiroyasu Yuhashi\,$^{1}$ and Teppei Koguchi,$^{1}$ }

\def\Address{$^{1}$Faculty of Informatics, Shizuoka University, Japan }
\def\corrAuthor{Junya Morita}

\def\corrEmail{j-morita@inf.shizuoka.ac.jp}

\begin{document}
\onecolumn

\title[Regulating Ruminative Web-browsing]{Regulating Ruminative Web-browsing Based on the Counterbalance Modeling Approach} 

\author[\firstAuthorLast ]{\Authors} %This field will be automatically populated
\author[\firstAuthorLast ]{\Authors} %This field will be automatically populated
\address{} %This field will be automatically populated
\correspondance{} %This field will be automatically populated

\extraAuth{}% If there are more than 1 corresponding author, comment this line and uncomment the next one.

\maketitle

\begin{abstract}

Even though the web environment facilitates daily life, emotional problems caused by its incompatibility with human cognition are becoming increasingly serious. To alleviate negative emotions during web use, we developed a browser extension that presents memorized product images to users, in the form of web advertisements. This system utilizes the cognitive architecture Adaptive Control of Thought-Rational (ACT-R) as a model of memory and emotion. A heart rate sensor modulates the ACT-R model parameters: The  emotional states of the model are synchronized or counterbalanced with the physiological state of the user. An experiment demonstrates that the counterbalance model suppresses negative ruminative web browsing. The authors claim that this approach is advantageous in terms of explainability.

\end{abstract}

\section{Introduction}
Even though the information provided by the web has drastically changed our lives and society, several emotional problems that had previously been rare have emerged. Such issues are currently becoming increasingly serious, occasionally inducing negative social collective behaviors such as flaming, cyberbullying, and cyberstalking. Even though social and legal systems for regulating these issues have been actively discussed (e.g.,  \citep{regulation2018general,jones2018ctrl+}), interventions based on the understanding of the underlying cognitive and emotional mechanisms remain underdeveloped. %Therefore, this paper focuses on behavior addiction and ruminative thinking as a fundamental emotional problem in the online society.

Concerning these mechanisms, it has been pointed out that the negative side-effect of technology is emphasized, as it is metaphorically as severe as drug addiction, which leads to repetitive overuse \citep{alter2017irresistible}. In this discussion, it was claimed that digital technology causes behavioral addiction because it enables users to effortlessly obtain information. Behavioral addiction is considered particularly problematic if users have mental health problems \citep{twenge2018decreases}. Rumination is a psychological state related to a depressive mood and involves repetitive negative thinking on a specific topic. Technologically enhanced and socially accumulated ruminative thinking may have a severe impact on society, with the synergistic action of the echo-chamber effect through a large number of individuals \citep{del2016echo,wollebaek2019anger}.

In this study, we propose an approach that combines cognitive modeling and information prompting as a nudge to prevent such negative behavior during web browsing. The proposed system is based on monitoring the physiological states of users. Specifically, we propose a counterbalance modeling approach to maintain an internal affective state following homeostatic regulation \citep{10.3389/fphys.2020.00200,cannon1929organization}. This paper is organized as follows. First, we review related work. Subsequently, we present the proposed approach and experimentally evaluate it. In the final section, we discuss implications including ethical issues and future directions.

\section{Related Studies}
Herein, we present a detailed background for this study: (a) problem statement, (b) user modeling, and (c) behavior change techniques.

\vspace{20pt}
\subsection{Ruminative Thinking and Behavioral Addiction}
This study is concerned with rumination, which is commonly defined as repetitive and negative thinking about unpleasant experiences, such as disappointments or past mistakes \citep{nolen1991prospective,treynor2003rumination}. Rumination is considered a serious mental health issue, and it has been claimed to be a preceding stage of depression. As is the case with other depressive symptoms, rumination prolongs dysphoric moods and causes attentional biases toward negative information  \citep{cramer2016major,mogg1995attentional}, that is, ruminating people are attracted to negative information more easily, resulting in worsened depressive symptoms. The level of severity varies, depending on the individual. In some cases, people who often ruminate face a greater risk of depression  \citep{kuehner1999responses}.

%Rumination is also considered as a type of task-unrelated thoughts \cite{christoff2016mind}, e.g., mind-wandering. This kind of mental states has the same characteristic of temporarily losing cognitive control of attention on the task on hand and deliberately engaging in contemplation -- continually recalling past experiences or imagining about the future. While mind-wandering could affect the task both positively (i.e., improving creativity; \cite{baird2012inspired}) and negatively (i.e., losing focus; \cite{mcvay2009conducting}), rumination which the memory retrieval process repetitively recalls unpleasant experiences causes only negative effects.

Combining such a negative mental state with information technology sometimes causes a more severe feedback loop accompanied with negative feelings. Information technology removes the limits from cognitive boundaries that have evolved throughout the history of humanity. Humans naturally forget information that is not relevant to their current \citep{anderson1991reflections,schacter2002seven}. However, the web can instantly provide Information that does not decay with time. This convenience induces addictive behavior, and people cannot resist using digital technology even when they are aware of the irrationality of their own behavior \citep{alter2017irresistible}.

%addictive internet use .
%The concept of Nuddge can be a possible solution.

%Targeting
\vspace{20pt}
\nocite{anderson2009can}
\subsection{Cognitive Modeling and Cognitive Architecture}
Considering that emotional problems in the web are caused by incompatibility with human memory, we focus on a user model that coordinates between natural and artificial cognitive systems. Among several approaches to modeling human cognition, the Adaptive Control Thought-Rational (ACT-R: Anderson, 2007) is selected in this study. It is one in a series of successively developed cognitive architectures \citep{Kotseruba2018}. A cognitive architecture provides basic modules of functions corresponding to human-brain regions. By utilizing these modules, it is possible to construct a cognitive model that simulates mental process occurring when a specific task is performed. The ACT-R architecture is especially suitable for constructing a model of the memory process of users during web browsing because it has {\it declarative modules} simulating the retrieval process of human declarative memory.

In fact, theorization of memory occupies a central position in ACT-R; it refers to the activation mechanism for each memory item, affecting the likelihood that the item will be retrieved successfully and quickly. The calculation of activation is largely controlled by the frequency and recency of each individual memory \citep{anderson1989human}. Frequently or recently retrieved memory items have high activation. In addition, memory decay and the spacing effect impact activation. The forgetting curve theory \citep{ebbinghaus1885gedachtnis} is concerned with memory decay, resulting in loss of information over time if this information is not recalled. By contrast, according to the spacing effect theory \citep{ebbinghaus2013memory}, frequently recalling information may strengthen its retention, that is, a piece of information becomes more difficult to forget after being periodically recalled.

Following such a memory activation mechanism, ACT-R naturally produces ruminative behavior; repetitive recalls of past experiences raises the priority of small number of memories, leading a continuous cycle of retrieving specific memories.
In fact, \cite{lebiere2009balancing} pointed out that a free recall made by the normal ACT-R model lead  ``pathological behaviors such as out-of-control looping. Moreover, during rumination, the brain continues retrieving memories based on their priority; nonetheless, the negative experiences that are the cause of rumination are the most likely to be recalled. 
Recent research has explored the simulation of such negative  memory retrieval using ACT-R. \cite{van2015modeling} constructed an ACT-R model that simulates the processes of mind-wandering. In this model, a state of attention on a task is suppressed by falling into the mind-wandering state in which the model continuously recalls past memories until it is reminded to return to its task. In another study \citep{van2018does}, the previous ACT-R model was applied, with moods (cheerful, content, down, suspicious, and insecure) in place of memories, and a model that simulates ruminating participants was implemented. The model supports the assumption that the activation of negative mood leads to the cyclic retrieval of highly prioritized memory, which is also consistent with an empirical study 
\citep{van2012effects}.

%\subsection{Model-based reminiscence}
In the present study, we apply previously obtained findings to ruminative behavior during web-browsing. The ACT-R has been used to support human activity in a series of studies. A well-known example can be found in education. Since its early success, the concept of an {\it intelligent tutoring system} has been greatly developed \citep{Anderson456}. However, this approach is only applicable to fields where the required knowledge is well-defined, such as mathematics, science, or programming. In contrast to this traditional approach, recent studies on ACT-R modeling have increasingly included affective processes \citep{dancy2015using,juvina2018modeling}. Based on such models, several methods of emotional support have been proposed. For example, \cite{morita2016model} developed a method called {\it model-based reminiscence}, in which the model contains user lifelog data as declarative memory, and presents these memory data in accordance with the retrieval mechanism implemented in ACT-R. This framework has been expanded by \cite{itabashi2020} to include physiological data such as heart rate so that the retrieval parameter may be modulated in real time. However, model-based reminiscence is limited because the system cannot intervene in ruminative web browsing in action.

\vspace{20pt}
\subsection{Prompting Behavior Change}
To support activity during web browsing, we should develop an information presentation method that does not strongly intervene in the main tasks. The concept of a \textit{nudge} refers to an approach realizing such an implicit intervention that mildly changes individual behaviors and decisions \citep{thaler2009nudge}. This concept itself includes various approaches, as reviewed by \cite{caraban201923}. Among the techniques categorized in the human–computer interaction (HCI) field, implicit intervention in human cognition has been found to have great potential to reinforce behaviors through drawing attention at appropriate times. For example, \cite{zhu2017exploring} studied the effects of implicit prompts on encouraging computer users to correct their sitting posture.
\cite{ibragimova2015smart} designed a smart driving system that collects data from an accelerometer to detect driver behavior and provides feedback through LED lights and vibrations on the steering wheel, nudging drivers to drive less aggressively.
Summarizing these studies, it is conceivable that nudge techniques (mild implicit prompting) could be extended to address mental issues, particularly rumination, and be applied to prevent users from continuing rumination during web browsing.

\section{Model-based Advertisements}
\begin{figure}[btp]
\begin{center}
 \includegraphics[scale=0.5,bb={0.000000 0.000000 895.000000 277.000000}]{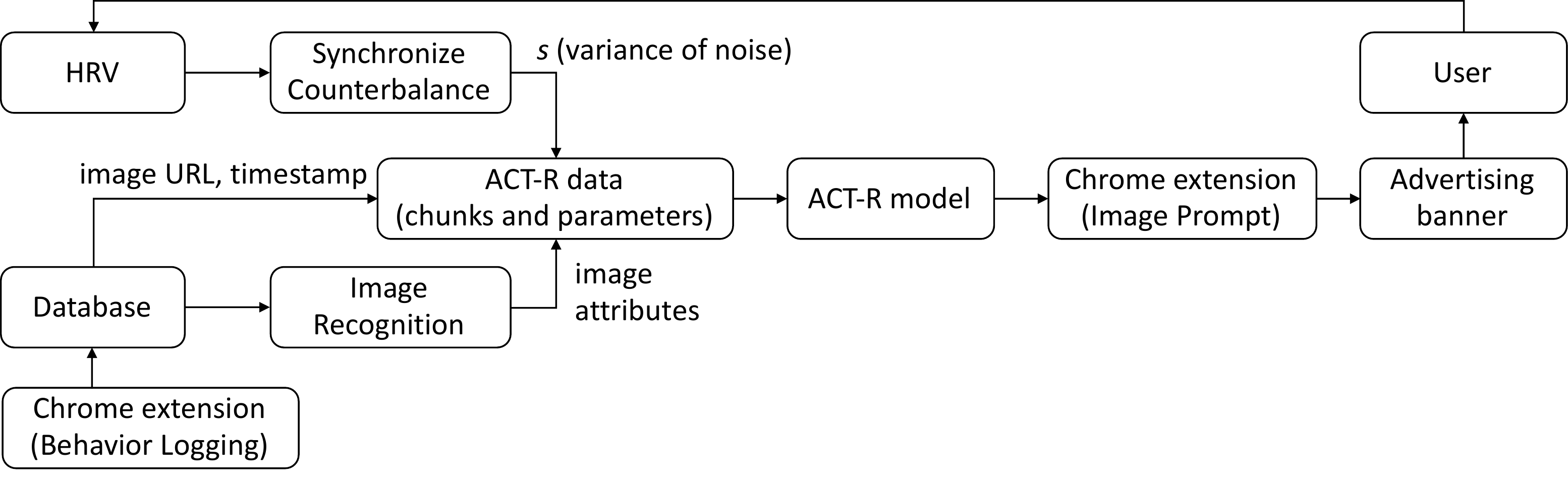}
%\vspace{-0.2in}    
 \caption{System overview.}
\label{fig:sys}
%\vspace{-0.2in}
\end{center}
\end{figure}

The authors extended the memory model developed in a previous study \citep{itabashi2020,morita2016model} to suppress ruminative behavior during web browsing. To naturally apply the memory model in a web environment, we focused on web advertisement such as behavioral targeting. Several studies have indicated the potential of behavioral changes toward healthy behavior through this type of online media \citep{Kramer8788,yom2018effectiveness}. In our system, the visited product images are always presented in the right region of a web page. The images are periodically changed to affect the implicit memory processes of the user. Figure \ref{fig:sys} shows an overview of the system. In the rest of this section, we explain the implementation of the system, which comprises three parts: (a) browser extensions, (b) a cognitive model, and (c) heart-rate sensing. 
The supplementary material in this paper also includes an example of actual behavior of the system as a movie.

\vspace{20pt}
\subsection{Browser Extensions}

The system is implemented in a web environment based on the widely used web browser Google Chrome. To customize the browsing experience on Chrome, Google provides a tool to develop ``extensions,'' built on web technologies (such as HTML, CSS, and JavaScript). Using this tool, we developed the following two independent extensions.

\vspace{20pt}
\subsubsection{Behavior Logging}
This extension collects user experiences in natural web-browsing activities. When the user visits specific shopping sites, such as Amazon, the extension automatically collects images (source URLs) displayed on the browser screen and stores them in a database on a server. The stored URLs are automatically converted into data according to the ACT-R declarative memory. Google cloud vision extracts image attributes, and the visited timestamp is also digitized by X-means clustering \citep{pelleg2000x}. These attributes (image tags and visited time) connect the observed images, forming a network that allows the model to successively retrieve images according to their semantic association \citep{morita2016model}. 

In addition to these symbolic attributes, each image holds numerical parameters: the number of times shown on the browser and the time since the images were shown. These parameters are used to simulate the frequency and recency effects of human memory \citep{anderson1989human}.

\vspace{20pt}
\subsubsection{Image Prompt}
The images stored in the database are retrieved by another Chrome extension that is connected with the ACT-R model. When this extension is enabled by the user, the model returns an image URL that is converted into an image by the browser. The ACT-R model and the Chrome extension communicate using TCP/IP, and the image in the right region of the browser is refreshed every 5 s according to the output of the model.

\vspace{20pt}
\subsection{Models}
We adopted the same ACT-R model as in the system used by \cite{itabashi2020} except that the present system employs product images rather than private photos. The model has declarative memory representing visited product images, and their retrieval is regulated using the standard activation equation of ACT-R 6.0 \citep{bothell}.
\begin{equation}
\label{activation}
  A_i = B_i + S_i + \epsilon_i
\end{equation}
The activation value ($A_i$) of the memory item $i$ is calculated as the sum of the base-level activation ($B_i$), the strength of association ($S_i$), and a probabilistic noise ($\varepsilon_i$). Among these factors, the base-level activation is crucial for representing the influence of an emotional state, and it is calculated as follows:
\begin{equation}
\label{b_i_nil}
  B_i = \ln(\textstyle\sum\limits_{j=1}^n t_j^{-d})
  +\beta_i
\end{equation}
where $n$ is the number of occurrences of memory item $i$, $t_j$ is the time elapsed since the $j$th occurrence, $d$ is the decay factor, and $\beta_i$ is the offset value. Thus, recently collected and frequently visited images receive high activation.

%Theoretical assumption behind this behavior is that human memory provides the most important information for the current time. However, such tendency leads pathological rumination.
With the base-level activation alone, the ACT-R memory naturally converges to a specific memory item, leading to repetitive displays on the browser \citep{lebiere2009balancing}. To escape this repetitive memory recall, the noise parameter $\varepsilon_i$ plays a crucial role. It is generated using a Gaussian distribution of mean 0, and variance
\begin{equation}
\label{noise}
\sigma=(\pi/3) \times s^2
\end{equation}
where $s$ is a parameter that determines the size of noise variance. If the value of $\varepsilon_i$ fluctuates widely, the possibility of escaping this feedback loop increases \citep{morita2016model}. To modulate  the noise level, \cite{dancy2015using} proposed a theory that connects the parameters of ACT-R to physiological indices, and explains the effect of emotion on cognitive processes. In their model, the parameter of noise variance ($s$ in Equation \eqref{noise}) is related to the activation of noradrenaline, which affects anxiety and arousal level \citep{mizuki1996differential}. That is, if a stressed situation is modeled, the output converges, whereas, in a relaxed situation, the model outputs a variety of behaviors.

Based on the theory by \cite{dancy2015using}, \cite{itabashi2020} developed a real-time parameter modulation method targeted at the current user. In this method, the noise parameters are replaced with the physiological state of the user, which is directly obtained from a heart-rate sensor. The present study also follows this idea, but two connection methods are employed, as described in the following subsection.

\vspace{20pt}
\subsection{Heart Rate Sensing}

%As we reviewed the previous research utilizing physiological data (heart rate data) in ACT-R model, this system also complies with the utilization of heart rate data of the previous study to calculate the noise parameter $\varepsilon_i$. However, in this study, the noise calculation method is different -- HRV is inverted before being converted to the parameter $\varepsilon_i$ unlike the previous model that HRV is directly converted.

The system collects physiological data from a heart-rate monitor and converts them into the noise parameter. Specifically, the R-R interval (RRI) of the user is constantly collected through a wearable heart-rate monitor (myBeat WHS-1, Union Tool Co. Japan), which is attached onto the chest of the user and is connected with the server  through Bluetooth. After the ACT-R model is activated, the program converts the data to $s$ in Equation \eqref{noise} in real time. This procedure follows the previous study by \cite{itabashi2020} and consists of the following three steps:

\vspace{25pt}

\subsubsection{Heart-rate Variability Calculation}

The heart-rate monitor collects and sends the RRI every three beats in the form of a single signal. The program first calculates the heart-rate variability (HRV) from the standard deviation of the three most recent  RRIs according to the following equation: 
\begin{equation}\label{hrv}
    HRV_i = \sqrt{(\sum(x_i - \mu)^2)/n}
\end{equation}
where $x$ is a single RRI; in fact, this is simply the standard deviation.

\vspace{20pt}
\subsubsection{Standardization}\label{stands}

The standard score of the HRV is calculated by comparing it with the mean of baseline data ($\mu$) collected when the user is in a relaxed state:
\begin{equation}\label{base}
    z = (HRV_i - \mu_b)/SD_b
\end{equation}
where $i$ represents the latest data, and $b$ the baseline data. Thus, this equation standardizes each HRV based on the variance obtained in a baseline session.

\vspace{20pt}
\subsubsection{Conversion of HRV to Noise Parameters}\label{conversion}
Finally, the standard score is converted to the parameter $s$ by adding an offset value of 0.5, and setting it to zero if it is below zero:
\begin{equation}
    s =
\begin{cases}
z + 0.5,    & \text{if } z + 0.5 > 0\\
0,          & \text{otherwise}
\end{cases}
\end{equation}

\vspace{20pt}
\subsubsection{Inverting HRV}
The parameter $s$ calculated by the above procedure leads to a large fluctuation in memory retrieval when the user is relaxed because the HRV increases in the parasympathetic dominance state \citep{7106477,6681457}. That is, when the user feels stressed and anxious, the activation value calculated by Equations (1) and (2) always outputs the same images, having a pronounced recency and frequency effect. Therefore, the model by Itabashi et al. (2020), where $s$ corresponds to the HRV, synchronizes ruminative behavior with the participants.

\begin{figure}[btp]
\begin{center}
 \includegraphics[scale=1,bb={0.000000 0.000000 378.000000 162.000000}]{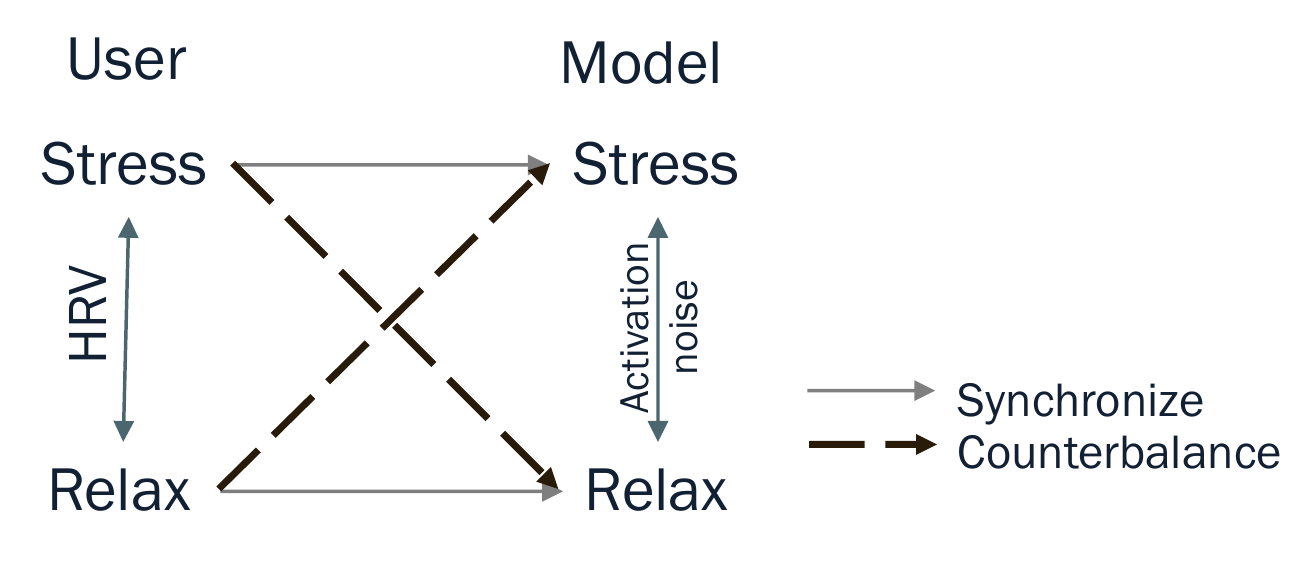}
%\vspace{-0.2in}    
 \caption{Connection with user and model.}
\label{fig:hom}
%\vspace{-0.2in}
\end{center}
\end{figure}

Unlike in the above method, in this study, it is hypothesized that the counterbalance model in which $s$ is the inverse of the HRV can aid in suppressing ruminative behavior. Figure \ref{fig:hom} shows the schematic representation of the model–user relation in the previous study (synchronization) and the present study (counterbalance). Accordingly, we modified the calculation. Specifically, after Equation \eqref{hrv}, we inserted an intermediate step by inverting the HRV and multiplying by the average of the baseline as follows:

\begin{equation}\label{inv}
    HRV_{inv} = HRV_i^{-1} \times \mu_b
\end{equation}

Using this $HRV_{inv}$, we proceeded with the above steps (\ref{stands}, \ref{conversion}). In this process, the baseline data of HRVs are also inverted.

\section{Experiment}
An experiment was conducted to test the efficacy of the proposed system at suppressing ruminative behavior. Achieving this requires inclining participants to ruminate. Therefore, we created a session to adjust the mood of the participants using mood induction procedures (MIPs), which are a psychological method to influence the emotions of the participants prior to the experiment \citep{kuvcera2012using}. Among several variations of MIPs, one common technique is \textit{imagination technique}, which evokes past emotional experiences by recalling such events. Under this condition, we investigated whether the  participants, with the aid of the proposed system, could recover from ruminative web browsing.

\vspace{20pt}
\subsection{Method}
%The following method was approved by the research ethics board of the university with which the authors are affiliated.

\vspace{20pt}
\subsubsection{Participants}
We recruited 12 Japanese participants (six males and six females) who reported that they often visit shopping websites (Amazon or Rakuten) through personal contact. All of them were undergraduate or graduate students majoring in informatics. They received a reward of 1,500 JPY for their participation.

\vspace{20pt}
\subsubsection{Design}

The participants were randomly divided into two groups (each comprising six people): synchronized (sync) and counterbalanced (coun). For the former, we used the ACT-R model with the $s$ parameter corresponding to the HRV (Equation \eqref{hrv}), whereas for the latter, we used the ACT-R model with $s$ being the inverse of the HRV (Equation \eqref{inv}). We hypothesized that the coun group would outperform the syn group in terms of suppressing ruminative web browsing.

%While the dependent variable in this study is the $s$ calculation method, we examined and compared the results of participants' evaluation about the degree of the negative mood between the groups through an analysis of variance (ANOVA), a statistical method. Therefore, the independent variable is the evaluation scores in the aspect of distraction.

%\subsubsection{ANS Settings}
%The $s$ calculation method of normal condition accords to equations equation 3 to equation 2 which the HRV is compared with baseline data to calculate the standard score instead of the recent 49 HRVs. In inverted ANS condition, we inserted between equation 1 and equation 2 meaning that HRV is inverted before compared with baseline data.

%We constructed a baseline session at the beginning of the experiment where the participants were told to sit back and relax in order to collect stable heartbeat data when being in a relaxing mood. Then in the main task, the HRV is compared to see how much it fluctuates from the baseline data. %If the HRV extremely differs, that can be said that the participant is in a stressful mood and having negative emotion. On the other hand, if the HRV is close to the baseline data, it means that the participant is relaxing. 
%The baseline data consists of 50 HRVs, and we utilized those from the 15th to 64th to avoid the transition of heartbeat data after wearing the device.

\vspace{20pt}
\subsubsection{Questionnaire}
To evaluate the effectiveness of the proposed system, the participants were asked four questions: 
\begin{enumerate}
    \item \textit{``How much did you recognize the images shown on the right region of the screen?''}
    \item \textit{``How much did you find interesting the images?''}
    \item \textit{``How much did you remember negative events from your memory?''}
    \item \textit{``How annoying did you find the images when the images were changed?''}
\end{enumerate}

Each question corresponds to an aspect presented in Table \ref{tab:q}, where three aspects (1, 2, 4) were adopted from a previous study that evaluated a prompt interface aimed at behavioral changes \citep{zhu2017exploring}, whereas the third item (distraction) was originally employed to evaluate the suppressing effect of the system on negative memory induced by a mood induction procedure that is described below. In the present study, all the aspects were rated on a seven point Likert-scale (1--7: very little to very much), except for the third aspect, for which a lower score indicates that the participants were distracted from negative memory retrieval to a larger degree.

\begin{table}[htb]
\caption{Aspects of post-experiment questions.}
\label{tab:q}
\centering
\begin{tabular}{|p{0.2\linewidth}|p{0.7\linewidth}|}
\hline
Aspect & Description\\
\hline
1) Recognition & The extent to which the participant can recognize the
images.\\
%\hline
2) Attention & The extent to which the participant is interested in the images.\\
%\hline
3) Distraction & The extent to which the images can distract the participant from negative memory retrieval. \\
%\hline
4) Annoyance & The extent to which the images annoy the participant.\\
\hline
\end{tabular}
\end{table}

\vspace{20pt}
\subsubsection{Procedure}
The experiment was conducted from June to July 2020 under the approval of Ethical Committee of Shizuoka University.
The procedure involved the following steps:

\begin{enumerate}

    \item Behavior logging: The participants downloaded and installed the extension for behavior logging on the computer they usually used. The duration from the download to the experiment was three to five weeks. During this period, the participants were asked to browse products on shopping websites.
    
    \item Instructions regarding the task: The objective and the entire procedure were explained to the participants. The aim was described as {\it ``examining the influence of memory recall by a system based on a cognitive model processing the user's web browsing history.''} Then, they were informed that the experiment included three tasks: {\it ``In the first and the second task, you will be asked to recall a recent memory and collect information about your future life, respectively. In the final task, you will write a report summarizing these two tasks.''} The final task was not actually carried out, but this instruction was made to maintain the influence of the mood induced by the first task on the second task. After the participants agreed to participate in the experiment, they were asked to sign a consent form.
    
    %\item The participant completes the pre-experiment questionnaire about mental conditions (DASS21).
    %\item We have the participant calibrated the eye tracker.
    \item Baseline measurement: Following the instructions, the participants attached the heart-rate monitor to their chest. After it was confirmed that the heart-beat data were transmitted correctly, the baseline session was initiated by letting all participants relax for approximately 3 min. The system utilized 50 HRVs (15--64th) obtained in this session to calculate the baseline in Equation \eqref{base}.
    
    \item Mood induction task: In this task, the participants recalled an unpleasant memory that they most frequently recalled in recent times. It could be any negative story that frequently came into their mind. Then, they noted it in a Microsoft Word document. The time limit was 15 min.
    
    \item Main task: In the main task, the participants were asked to assume that {\it ``they are going to start a new working life anywhere you want next spring.''} They were told to spend 15 min browsing websites and searching for what they thought would be necessary for realizing their plans. The content of the choices of the participants could be interpreted broadly. We assumed that participants who were unable to stop recalling negative experiences would be likely to ruminate in this type of open-ended question. During the task, the image-prompt system operated to display product images selected by the ACT-R model.
    \item Questionnaire: After the time was up, we explained that the summary task was a dummy. We finished the experiment by having the participants complete the post-experiment questionnaire presented in Table \ref{tab:q}.
\end{enumerate}

\subsection{Results}

The data obtained in the experiment were analyzed to study the effect of the counterbalance approach on the recovery process from mood induction. Owing to the limited number of participants, nonparametric statistical methods were adopted. Specifically, we used the Wilcoxon rank-sum test to examine the difference between the two groups. In addition, we considered results with $p < .10$ to be marginally significant, setting the significance level to $p < .05$.

\subsubsection{Model Behaviors}
\begin{figure}[tb]
\centering
  \includegraphics[width=\columnwidth,bb={0.000000 0.000000 594.000000 197.000000}]{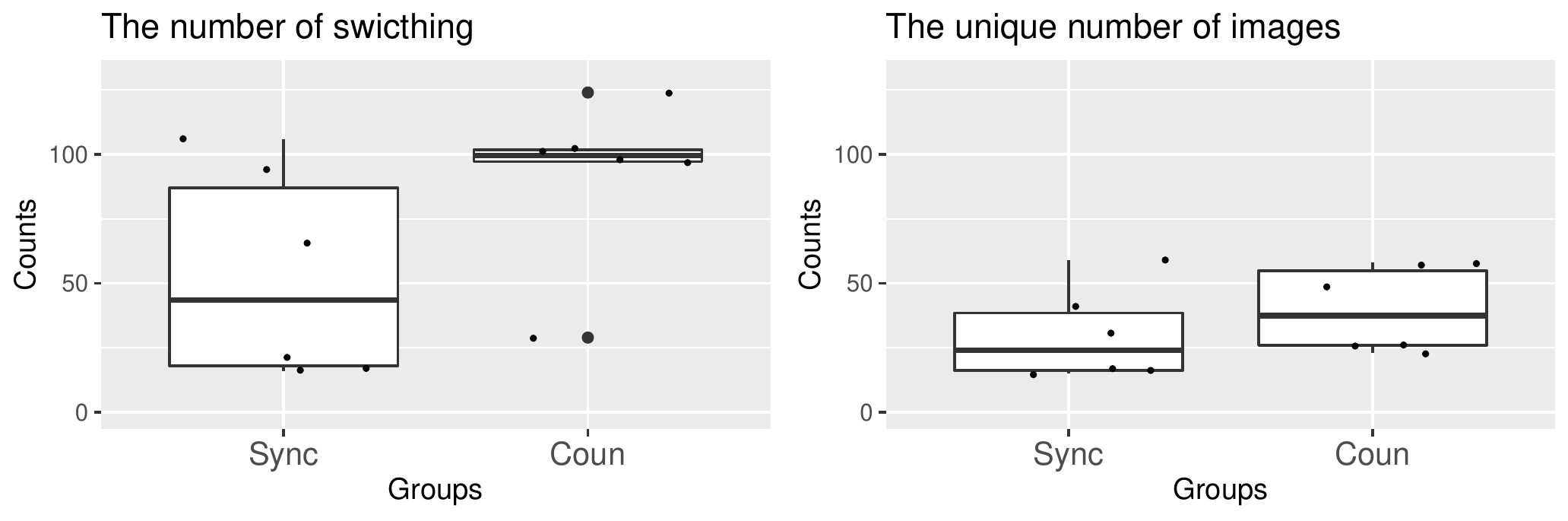}
\caption{Counting results for model behavior.}
\label{mb}
\end{figure}

As explained previously, the system refreshes the browser images every 5 s. However, owing to the recency and frequency effect introduced by Equation \eqref{b_i_nil}, the same images tend to be shown on the display. Therefore, we examined the effect of the counterbalance model by counting the total number of image switches and the unique number of images, excluding duplicates. The counts are shown in Figure \ref{mb}, where a marginally significant difference in the total number of switches between the two groups can be observed ($p = 0.09$). By contrast, the difference in the unique number of images is not significant ($p = 0.37$). These results indicate that the two models appear to behave differently although the difference is not obvious in these simple indices.

\subsubsection{Questionnaire Answers}
\begin{figure}[tb]
\centering
\includegraphics[bb={0.000000 0.000000 473.000000 219.000000},width=0.9\linewidth]{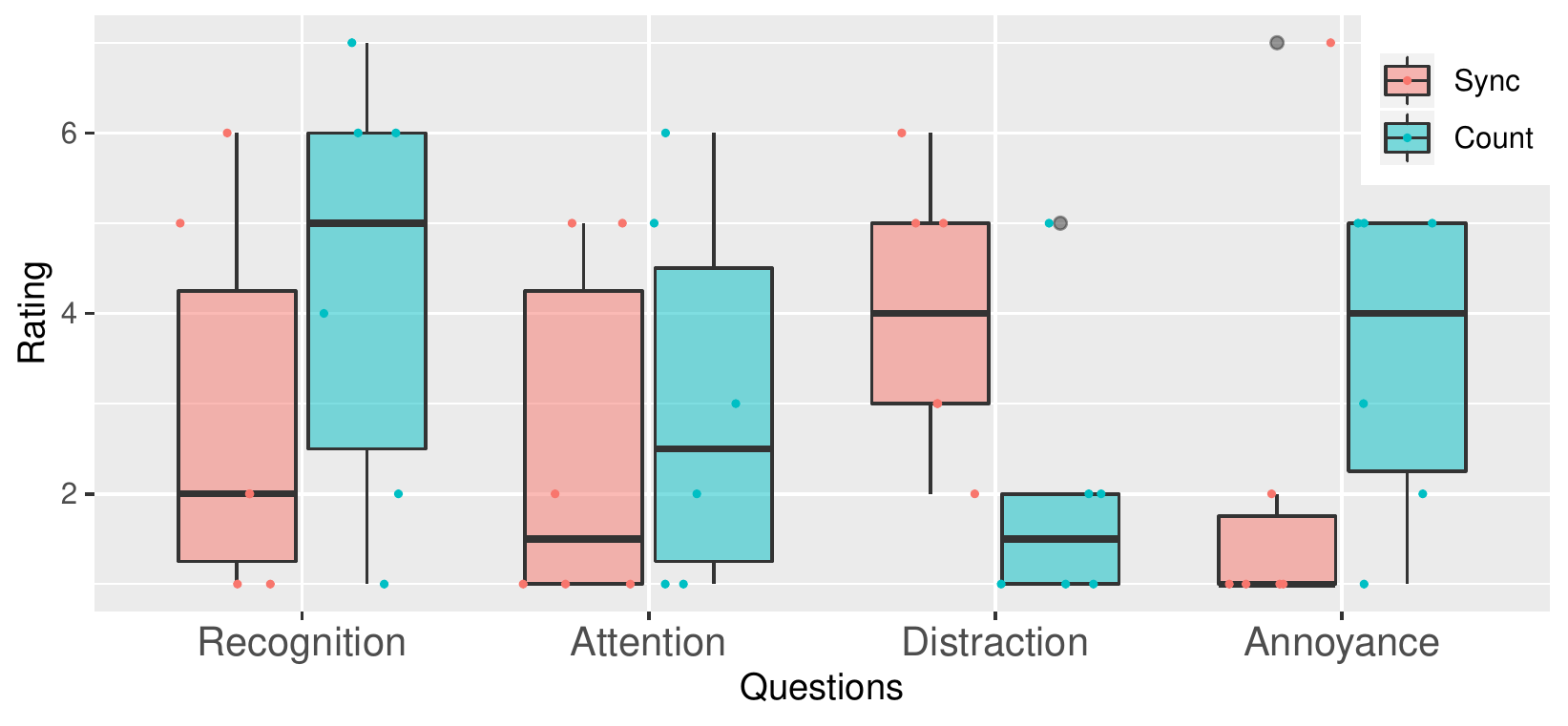}
\caption[Questionnaire answers]{The evaluation scores for the post-experiment questions.}
\label{fig:answers}
\end{figure}

To study the subjective effect of the system, we compared the scores obtained in the questionnaire, as shown in Figure \ref{fig:answers}. Among the four questions, a significant difference was found regarding distraction ($p = 0.04$). Accordingly, the effect of the proposed approach on suppressing negative memory retrieval was confirmed.
No significant difference between the two groups was observed regarding the other aspects (recognition: $p = 0.65$; attention: $p = 0.65$; annoyance $p = 0.180$), suggesting that the effect of the proposed approach on the main task was not so explicit.

\vspace{20pt}
\subsubsection{Heart Rate Data}
The graph on the left of Figure 5 shows the boxplots of the HRV calculated by Equation (4), which were summarized from Figure 6 and indicate the fluctuations in the HRV between the two tasks. The boxplot indicates the differences between the two tasks in both groups: The HRV in the mood-induction task was smaller than that in the main task. This may be because the negative memory induced by the mood induction task might lead to a small HRV, or the participants were nervous at the start of the experiment. For whatever reason, in the main task, the participants recovered their mood while browsing for their future life.

\begin{figure}[tb]
\centering
  \includegraphics[bb={0.000000 0.000000 592.000000 265.000000},width=\columnwidth]{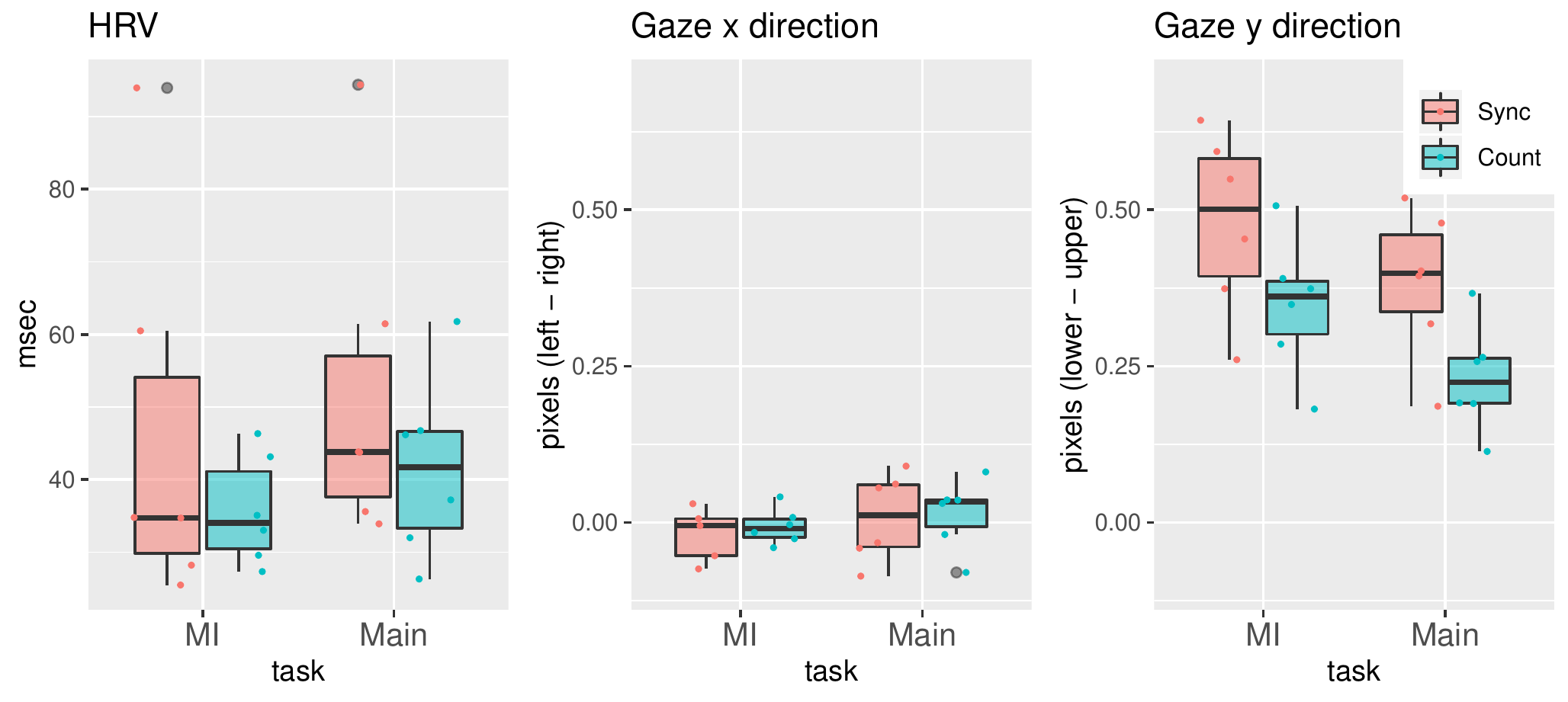}
\caption{HRV and gaze distributions.}
\label{fig:hrdiff}
\end{figure}

\if0
\begin{figure}[tb]
\centering
\begin{subfigure}{0.45\columnwidth}
    \centering
    \includegraphics[width=\linewidth]{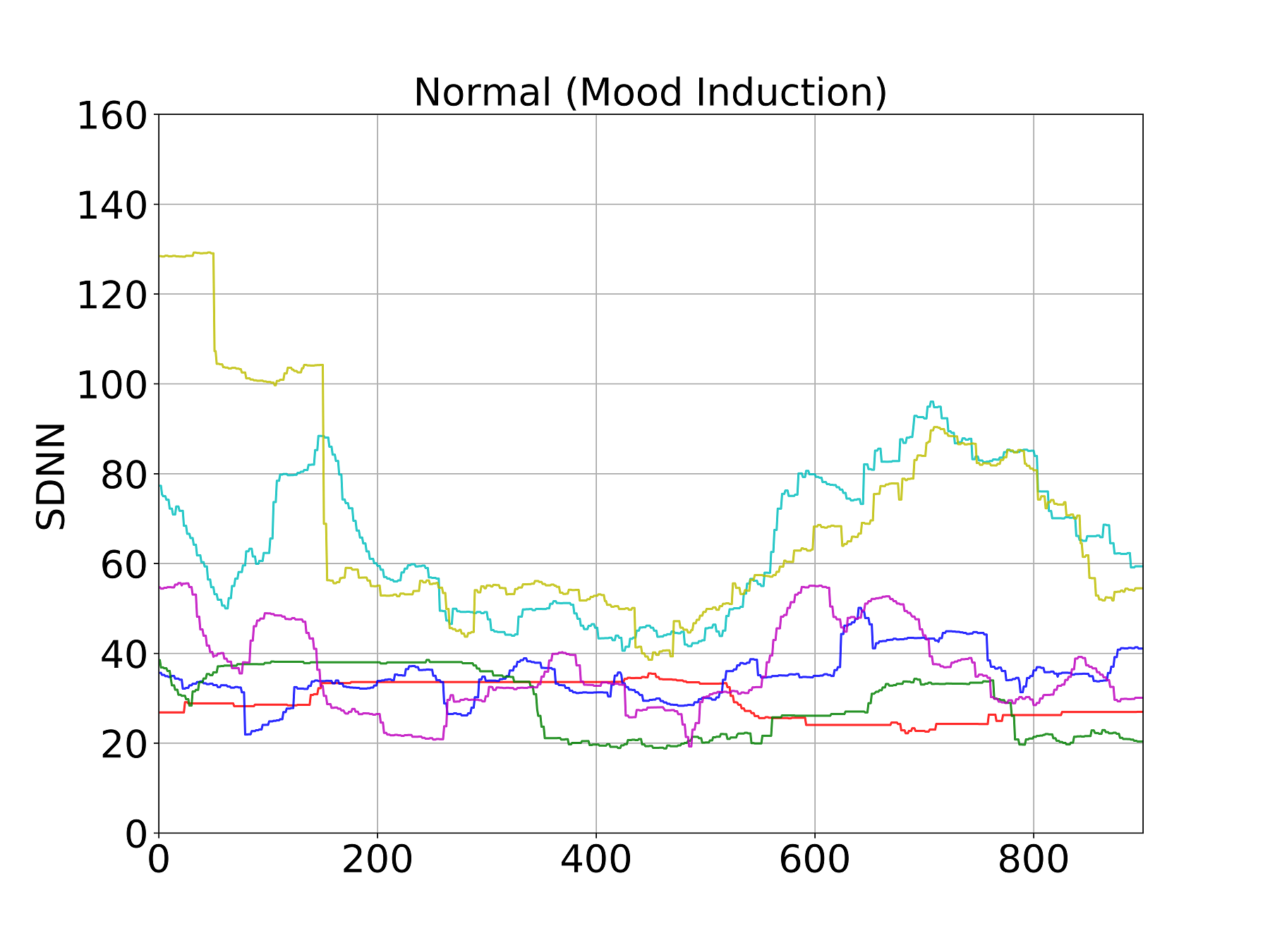}

\end{subfigure}
\begin{subfigure}{0.45\columnwidth}
    \centering
    \includegraphics[width=\linewidth]{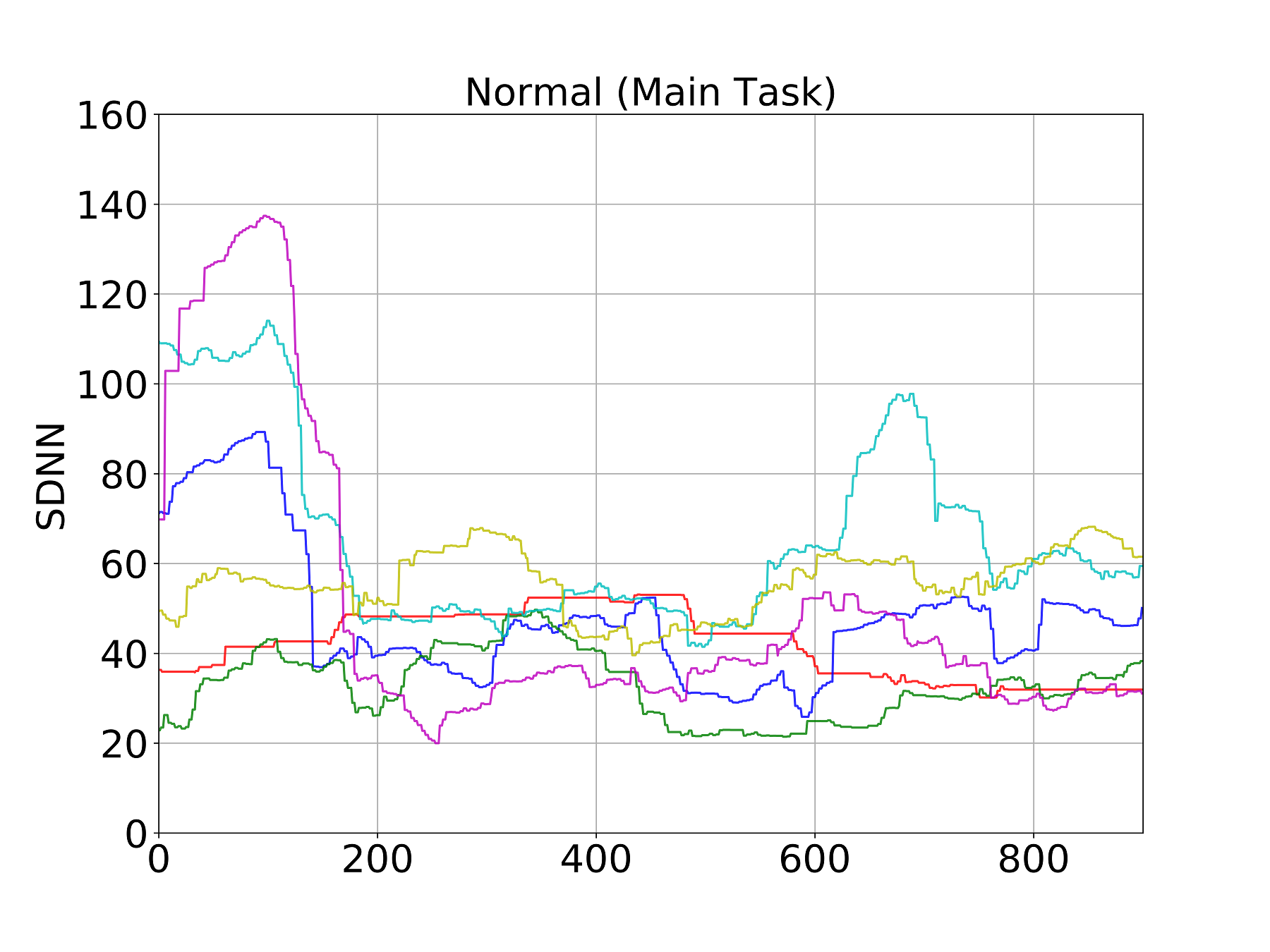}
\end{subfigure}
\bigskip
\begin{subfigure}{0.45\columnwidth}
    \centering
        
    \includegraphics[width=\linewidth]{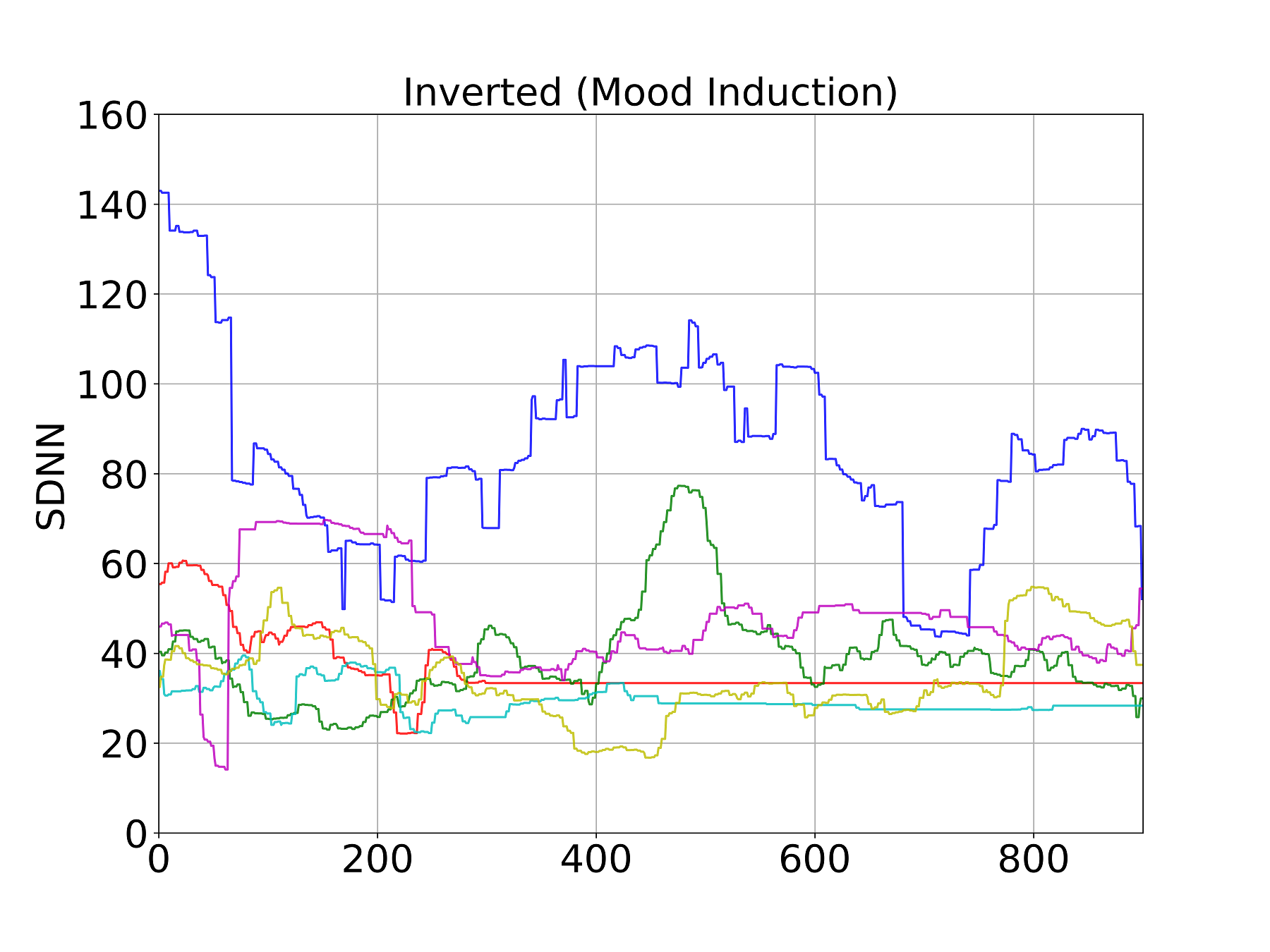}

\end{subfigure}%
\begin{subfigure}{0.45\columnwidth}
    \centering
    \includegraphics[width=\linewidth]{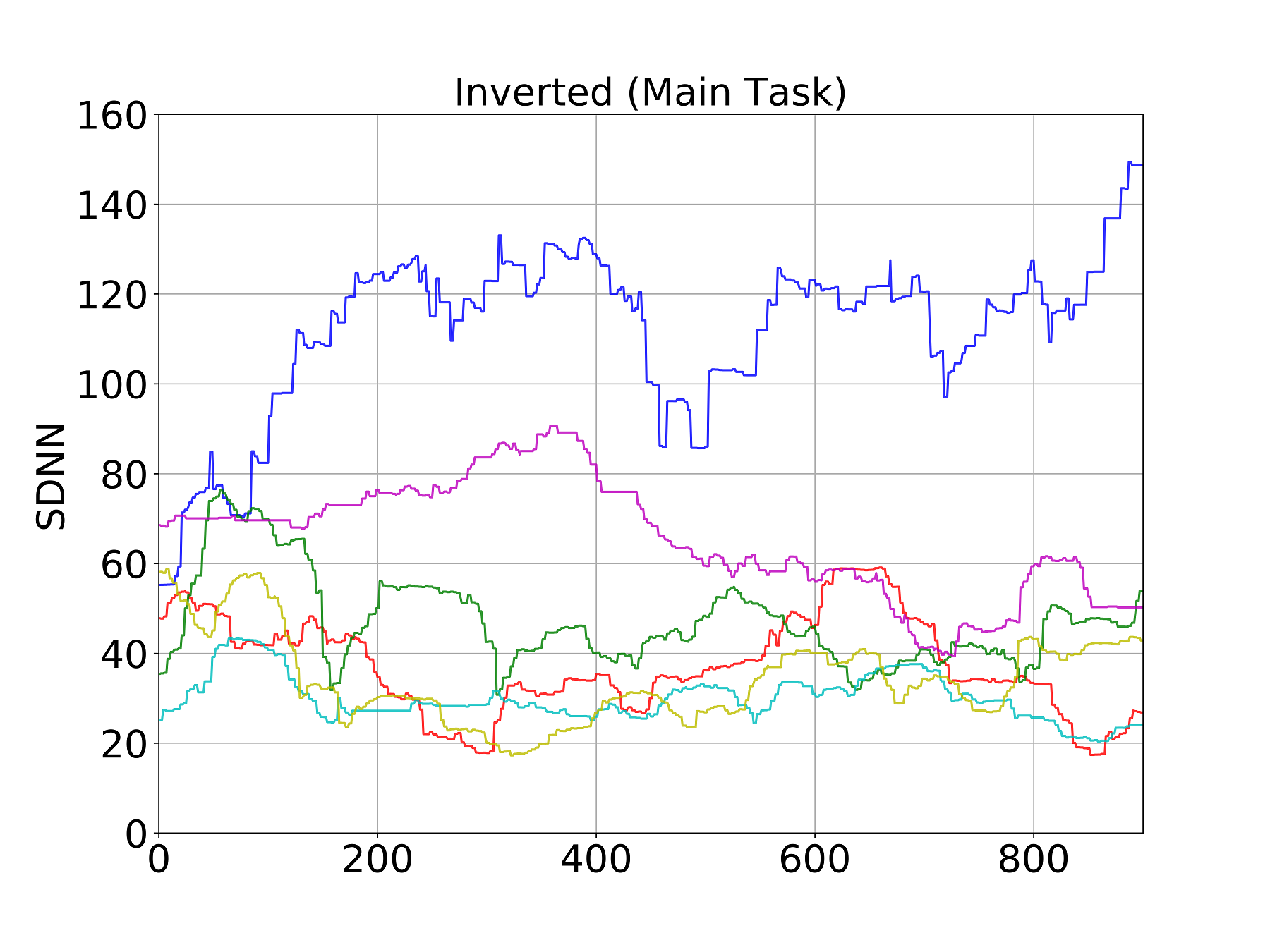}
\end{subfigure}%
%\bigskip
%    \begin{subfigure}{1.2\columnwidth}
%    \centering
%    \includegraphics[width=\linewidth]{Participant legends.png}
%\end{subfigure}
\caption{Fluctuations of HRVs during the two tasks (left: the mood induction, right: the main task). Each line indicates the participant in the two groups (top: synch, bottom: count).}
\label{fig:rri summary}
\end{figure}
\fi

\begin{figure}[tb]
\centering
  \includegraphics[bb={0.000000 0.000000 893.045893 687.035306},width=0.9\columnwidth]{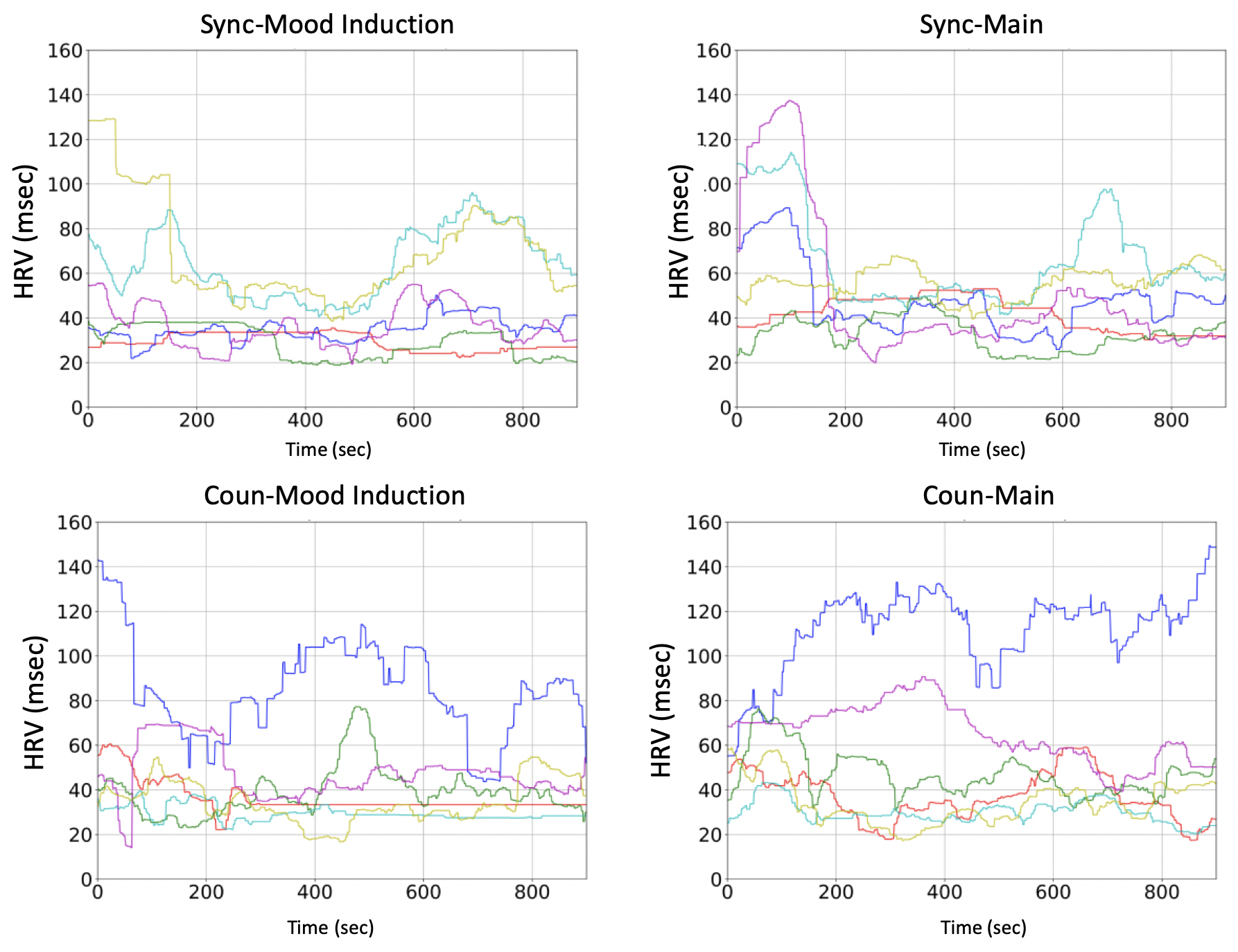}
\caption{Fluctuations of HRV during the two tasks (left: mood induction, right: main task). Each line indicates a participant in the two groups (top: sync, bottom: coun).}
\label{fig:hrfl}
\end{figure}

Unlike the difference between the two tasks, the difference between the two groups in terms of the recovering effect was not obvious. We tested the difference between the groups by comparing the difference of the HRV from the main task to the mood induction task. No significant difference between the groups was observed regarding this index (median in sync: 7.90, median in coun: 2.26, $p=0.81$).

\vspace{20pt}
\subsubsection{Gaze Information}
We examined the difference in the gaze information during the tasks. As the interface used in this study presents images on the screen, we assumed that the influence of the system appeared in the gaze movement. In this analysis, we used the OpenFace software package \citep{8373812} to extract gaze information from a video recorded during the tasks. In the experiment, the faces of the participants were recorded with a web camera (1080p, 30 fps) set on the top of the display. Among several indices output from the OpenFace, we used the $x$ and $y$ values of the vectors directed from the eyes on the image captured by the camera. 

\begin{figure}[tb!]
\centering
    \includegraphics[bb={0.000000 0.000000 776.000000 508.000000},width=1\linewidth]{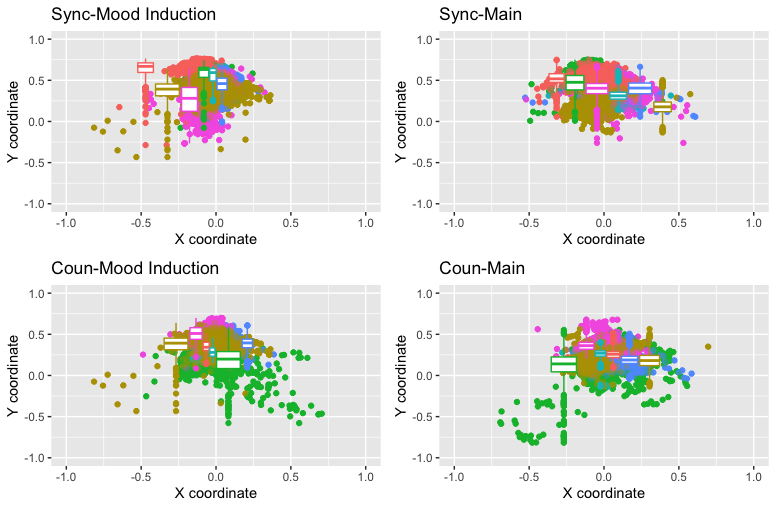}
\caption{Gaze distribution in the two tasks (left: mood induction, right: main task). The colors identify the participants in the two groups (top: sync, bottom: coun).}
\label{fig:gazes}
\end{figure}

Figure \ref{fig:gazes} shows the gaze distribution of the participants for each task in each group, presenting the $x$–$y$ coordinates as dots obtained each second (averaging 30 frames per second). The dot color distinguishes the participants, and the boxplots in the figure indicate summaries of the dot distribution for each participant. Although there were large individual differences between the participants in terms of this measure, we can find the difference between tasks in both groups. To clarify this difference, we further summarized the gaze distribution for each participant, as shown in the two panels on the right of Figure \ref{fig:hrdiff}, where each dot corresponds to the averaged $x$-$y$ coordinates for each participant. In the main task, the vectors from the eyes were directed rightward lower than in the mood induction task, suggesting the participants mainly observed the left part of the screen (the contents of the web page) in the main task. However, the difference between the two groups in terms of this tendency is less clear. Again, we calculated the difference from the main task to the mood-induction for each $x$ and $y$ coordinate. No significant difference was found for either the horizontal (median in sync: $0.04$, median in coun: $0.03$; $p = 0.58$) or the vertical direction (median in sync: 0.11, median in coun: $0.11$; $p = 1.00$) in this index.

\vspace{30pt}
\subsubsection{Correlation Analysis}
A significant difference between the two groups was found only in the evaluation of distraction. To confirm the reliability of this difference, we calculated Spearman's rank correlation coefficients between the distraction rating and the other indices. Table \ref{tab:corr} summarizes the results of this calculation for each group. Only a significant correlation in the number of switches in the counterbalance model ($p = 0.03$) can be observed. This suggests that the increase in switching by the counterbalance model is correlated with high negative memory retrieval. The horizontal gaze direction also showed a marginally significant correlation with distracting negative memory retrieval in the sync ($p = 0.05$) and coun groups ($p = 0.09$), although the direction of the correlation was different. In the sync group, attention toward the web contents (the left screen) is corelated with distraction of negative memory retrieval, whereas the attention toward the web contents in the coun group is corelated with the negative memory retrieval. In other words, in the coun group, the distraction of the web contents is correlated with the the distraction of the negative memory retrieval, suggesting the effect of the web advertisement.

\begin{table}[tb]
\caption{Spearman's correlation coefficient between the distraction and other indices. $+: p<0.10 *: p<0.05$}
\label{tab:corr}
\centering
\begin{tabular}{p{0.5\columnwidth}rr}
\hline
 & Sych&Coun\\\hline
Recognition & $-0.060$&$-0.547$\\
Attention &$0.127$ &$-0.046$\\
Annoyance &$-0.069$&$-0.557$\\
\hline
The total number of switches & $-0.117$&$0.833\rlap{*}$\\
The unique number of images& $-0.029$&$-0.172$\\
\hline
Difference of x-gaze from main to MI&$-0.794\rlap{+}$&$0.740\rlap{+}$\\
Difference of y-gaze from main to MI &$-0.588$&$-0.617$\\
Difference of HRV from main to MI& $-0.088$&$-0.308$\\
\hline
\end{tabular}
\end{table}

\section{Discussion and Conclusion}
\subsection{Summary and Implications}
To regulate ruminative web browsing, we developed a system that consists of a cognitive model of memory using ACT-R, physiological sensing to modulate memory retrieval, and image prompts implemented as a browser extension. This combination is considered to follow the principles of affective computing \citep{picard1995affective,picard2003affective}, which emphasizes computational models of affect and emotion, emotion recognition using multimodal data, and emotion expression affecting the mental state of users.

The contribution of the present study is to extend affective computing by including a computational cognitive modeling of memory. This model differs from the previous behavioral model \citep{10.1145/1541948.1541999} used in web advertisement \citep{yom2018effectiveness} in that it includes internal memory processes. Although there are several options for modeling emotion and memory \citep{friston2010free,schmidhuber2010formal}, we claim that including the ACT-R cognitive architecture provides another theoretical basis of implicit-prompting systems designed to adapt the emotional states of users based on an academic field with a long history.

%Grounding such a public knowledge, the design of the system is explainable in a wider community.

Furthermore, we empirically confirmed the effect of model-based prompts in the condition where the model maintains user homeostasis. The effect of this counterbalance approach was demonstrated in the subjective evaluation. The correlation analysis also validated this evaluation, indicating correlations with the number of switches and with the gaze directions toward the advertisements. These correlations confirmed that the proposed mechanism is effective in distracting negative memory retrieval. Thus, we can summarize the experiment by observing the recovery process in subsequent web-searching: the effect of the approach was observed in forgetting negative memories of users. Accordingly, the proposed approach can be seen as a possible candidate for achieving harmony between natural and artificial cognitive systems, balancing emotional issues in this digital age.

%It can be considered that the negative emotion during the web use is a common problem, leading to a debate of restricting a technology development. We consider this approach does not directly resolve such problem, but enhancing self-regulation during the web use help to suppress problematic behavior in the web.

\vspace{20pt}
\subsection{Limitations and Future Works}

Despite its advantages, the proposed approach has several limitations that should be addressed. Specifically, there is a lack of strong evidence for the effectiveness of the counterbalance model in terms of behavioral and physiological indices. The statistical tests in Figure \ref{fig:hrdiff} could not indicate significant effects demonstrating the advantages of the model. As can be seen in Figures \ref{fig:hrfl} and \ref{fig:gazes}, the variance from individual differences is so large that the number of participants should be increased. The large cost of  the experiment (e.g., long preparation period and large amount of physiological data) is the one of the reasons for the limited number of participants in the present study. The other reason is that the experiment was conducted amidst the worldwide COVID-19 pandemic, where activities were spontaneously suppressed. Although the small number of the participants is the limitation we should overcome in the future, we consider that the experiment conducted at such situation, where a high level of anxiety is expected, is also worth reporting.

Another limitation is the experimental setting under which the negative mood was induced. In such a situation, the counterbalance model frequently changed images to distract attention from negative memories. To fully utilize the counterbalance modeling approach, we should set other experimental settings, such as a situation in which an optimistic mood is induced. By inducing a relaxed mood, users' HRV becomes high and the counterbalance model presents fewer images. It is an interesting question whether such a situation promotes focus on web searching.

The limitation of the experimental settings in this study also questions the need for adaptive modulation using HRV. That is, some readers may find it sufficient to simply present images frequently without a parameter modulation by HRV. 
However, concerning this question, the present study alone can claim that the frequency of image changing without monitoring user's state is not sufficient to suppress negative memory retrieval. The correlation analysis presented in Table \ref{tab:corr} indicates non-monotonic relations between the model behavior and the suppression of negative memory retrieval. Focusing on the coun group, we observe that frequent switching rather promoted a negative memory retrieval. This suggests the advantage of adaptive modulation to maintain an optimal level of arousal \citep{yerkes1908relation}. In the future study, we will further explore methods of setting such an optimal level of arousal.

Finally, in the future study, we should analyze searching behaviors in the main task. Although this research targeted ruminative behavior, we did not present the occurrence of such behavior. The written text in the mood induction task included the death of pets, job hunting failures, and worries about interpersonal relationships. Some participants appeared to reflect the nervousness induced in the respective task in their web searches in the main task. For example, a participant who noted regret regarding his/her decision to attend graduate school also performed a search related to financial issues in the lives of graduate students in the main task. Another participant also used search keywords related to the remote work environment, reflecting the pandemic situation at the time. However, it was difficult to confirm the severity of such ruminative behavior from the existing data alone. In a future study, we will develop an analysis method to quantify ruminative behavior during web browsing.

\vspace{20pt}
\subsection{Ethical Stance}

As a final remark in this paper, the ethical aspects of this study should be stated. The study was conducted to prevent ruminative web browsing based on the belief that this personal behavior relates to social problems. If most people in cyberspace could regulate their behavior at will, many social disputes (e.g., Jones, 2018) would not occur. However, implicit prompts such as advertisements have always been a problem, as one can control the emotions of others without their consent. In this regard, the cognitive model-based approach may be a better choice because cognitive modeling has explicit parameters with shared consensus in academic communities, and therefore explainability is higher than in machine-learning user modeling.

We also believe that the methods of parameter modulation such as counterbalancing or synchronizing in this study should be selected by users. Even people with severe depression should be allowed to select how their behavior should be regulated. There is a moment when they can manage to make future plans \citep{10.1145/3313831.3376309}. Therefore, we believe that open and clear discussion made in academic communities will eventually overcome the problem caused by the technology developed in the community itself.

%\bibliography{refarence}

\end{document}